\documentclass[aps,prl,twocolumn,superscriptaddress,showpacs]{revtex4}

\usepackage[latin1]{inputenc}
\newcommand{\average}[1]{\left<{#1}\right>}

\begin{document}

\title{Comment on: Failure of the work-Hamiltonian connection
for free energy calculations}

\author{A. Imparato}
\email{alberto.imparato@polito.it}

\affiliation{Dipartimento di Fisica and CNISM, Politecnico di Torino,\\
Corso Duca degli Abruzzi 24, 10121 Torino (Italy)\\
INFN, Sezione di Torino, Torino (Italy)}

\author{L. Peliti}
\email{peliti@na.infn.it}

\affiliation{Dipartimento di Scienze Fisiche and CNISM, Università ``Federico II''\\
Complesso Monte S. Angelo, 80126 Napoli (Italy)\\
INFN, Sezione di Napoli, Napoli (Italy)}

\date{June 8, 2007}

\begin{abstract}
We argue that the apparent failure of the work-Hamiltonian connection
for free energy calculations reported by Vilar and Rubí 
(cond-mat arXiv:0704.0761v2) stems from their incorrect expression
for the work.
\end{abstract}


\maketitle
In a recent preprint, Vilar and Rubí~\cite{VR} have argued  that the pivotal
connection between the microscopic work $W$ performed by
a time-dependent force and the energy change associated with
its change in state cannot be used to estimate free-energy
changes. Since in several occasions some groups have exploited
the measure of work applied on microscopic systems to
evaluate its free energy landscape, the failure of this connection
would have far-reaching consequences.

In the present note we reevaluate Vilar and Rubí's arguments
to test their validity. We find that they rest on an incorrect 
expression of the work to be evaluated, and that therefore, if one
measures the correct quantity, which is experimentally accessible,
it is possible to exploit the work-free energy relation in order
to gather thermodynamical information on microscopic systems.

Let us consider a system whose microscopic state is described
by the coordinate $x$, and governed by the hamiltonian $H(x,\mu)$
which depend on an external parameter $\mu$ which can be manipulated. 
The equilibrium state for a given value of $\mu$ is described by
the canonical distribution
\begin{equation}\label{canonical:eq}
	p^\mathrm{eq}_\mu(x)=\frac{e^{-H(x,\mu)/T}}{Z_\mu},
\end{equation}
where the partition function $Z_\mu$ is given by
\begin{equation}
	Z_\mu=\int d x\;e^{-H(x,\mu)/T}.
\end{equation}
The free energy $G_\mu$ associated with the given value of $\mu$ is
given by
\begin{equation}\label{g2z:eq}
	G_\mu=-T \log Z_\mu.
\end{equation}

Let us now evaluate the change in $G_\mu$ as $\mu$ varies from, say,
$\mu=0$ to a final value $\mu$. One has
\begin{eqnarray}
	\frac{\partial G_\mu}{\partial \mu}&=&(-T)
	\frac{1}{Z_\mu}\int d x\;e^{-H(x,\mu)/T}\,
	 \left(-\frac{1}{T}\frac{\partial H(x,\mu)}{\partial \mu}\right)
	 \nonumber\\
	&=&\average{\frac{\partial H}{\partial \mu}}_\mu,
\end{eqnarray}
where $\average{A}_\mu$ denotes the average of the function
$A(x)$ with respect to the canonical distribution (\ref{canonical:eq}).
Thus
\begin{equation}
	\Delta G=G_\mu -G_0=\int_0^\mu d\mu'\;
	\average{\frac{\partial H}{\partial \mu}}_{\mu'}.
\end{equation}
We now define the \textit{reversible work} $W_\mathrm{rev}$
as the quantity which
appears on the rhs of this expression:
\begin{equation}
	W_\mathrm{rev}=\int_0^\mu d W_\mathrm{rev}=\int_0^\mu d\mu'\;
	\average{\frac{\partial H}{\partial \mu}}_{\mu'}.
\end{equation}
This definition is an immediate consequence of equation~(128.1), p.~535 of
Tolman's book~\cite{Tolman}. 
Its contents is explained by Tolman~\cite[p.~527]{Tolman} in the following way:
\begin{quote}
  The concept of \textit{work} performed by a thermodynamic
  system on its surroundings depends on the possibility of changing
  the values of external parameters for the system, which have the 
  nature of generalized coordinates. In the thermodynamic
  treatment, when a small variation is made in the value of such an external
  coordinate, the work done can be set equal to the product
  of that variation by the corresponding generalized force exerted
  by the system on its surroundings, and can be regarded as equal to the
  (potential) energy thereby transferred to external bodies. [\dots]
  In the corresponding statistical treatment we can assign to all 
  the systems in the representative ensemble the same value
  of the external coordinates as those of the thermodynamic
  system of interest, and can represent the corresponding
  generalized forces by their average values in that ensemble.
  The work done and the energy thereby transferred to
  external bodies can then be represented by its average value
  for the ensemble, as given by the product of the definite value of the
  displacement with the average value of the corresponding generalized force.
\end{quote}
With this definition, the equality between $\Delta G$ and
$W_\mathrm{rev}$ is trivially satisfied:
\begin{equation}\label{connection:eq}
	\Delta G=W_\mathrm{rev}.
\end{equation}
It is important to remark that, \textit{by definition}, the reversible
work is a non-fluctuating quantity (the equilibrium average of an
observable). Thus, if we evaluate, e.g., $\average{e^{-W_\mathrm{rev}/T}}$,
we have
\begin{equation}
  \average{e^{-W_\mathrm{rev}/T}}=e^{-W_\mathrm{rev}/T}
  =\exp\left\{\int_0^\mu d\mu'\;
  \average{\frac{\partial H}{\partial \mu}}_{\mu'}\right\}.
\end{equation}
On the other hand, because of (\ref{connection:eq}), one has
\begin{equation}\label{jarzrev:eq}
	\average{e^{-W_\mathrm{rev}/T}}=e^{-\Delta G/T},
\end{equation}
and, because of (\ref{g2z:eq}),
\begin{equation}
	e^{-\Delta G/T}=e^{-(G_\mu-G_0)/T}=\frac{Z_\mu}{Z_0}.
\end{equation}
Equation (\ref{jarzrev:eq}) represents 
the special case of Jarzynski's equality~\cite{Jarzynski} which
holds in the limit of reversible manipulation. Let us define the fluctuating
infinitesimal work, $d W(x,\mu)$ as the quantity whose average is
given by $d W_\mathrm{rev}(\mu)$:
\begin{equation}
  d W(x,\mu)=d\mu\,\frac{\partial H(x,\mu)}{\partial \mu}.
\end{equation}
Then the Jarzynsky equality states that
\begin{equation}
  \average{e^{-W/T}}=e^{-\Delta G/T}.
\end{equation}
In this expression, $W$ is the fluctuating total work
\begin{equation}
  W=\int d W =\int_{t'=0}^{t'=t} dt'\;\dot\mu(t')\,
  \left.\frac{\partial H(x,\mu)}{\partial \mu}\right|_{x=x(t'),\mu=\mu(t')},
\end{equation}
and the average is taken over all realizations of the process $x(t)$
with a given manipulation protocol $\mu(t)$.

In order to give a more definite interpretation of this quantity,
let us consider, as in~\cite{VR}, the simple case of a particle
bound to the origin by a spring of Hooke's constant $k$ and to
which a time-dependent force $f(t)$ is applied. 
We then take $f$ as a generalized coordinate, and write $H(x,f)$ as
\begin{equation}
  H(x,f)=\frac{1}{2}k x^2 -f x.
\end{equation}
Then
\begin{equation}
  \frac{\partial H}{\partial f}=-x,
\end{equation}
and
\begin{equation}
  d W=-x\,df.
\end{equation}
On the other hand, the \textit{mechanical} work that the external force
exerts on the system is given by
\begin{equation}
  d W_\mathrm{mech}=f \,dx.
\end{equation}
Note that
\begin{equation}
  dW-dW_{\mathrm{mech}}=-d(f x).
\end{equation}
We see that the work appearing in Jarzynski's equality is not equal
to the mechanical work exerted on the system. The difference
between the two works is equal to the variation of the potential
energy associated with the force. It is therefore no surprise
that, if one evaluates the integral of the mechanical rather than
the thermodynamical work, one obtains an incorrect estimate of
the free-energy difference.

We can now straightforwardly check the validity of the relation
between the reversible work and the free energy, provided the
full Hamiltonian is taken into account. We have, of course,
\begin{equation}
  d W_\mathrm{rev}=-df \,\average{x}_f,
\end{equation}
where $\average{x}_f$ is the equilibrium position of the particle
at force $f$:
\begin{equation}
  \average{x}_f=\frac{f}{k}.
\end{equation}
Thus
\begin{equation}
  \Delta G=-\int df\;\average{x}_f=-\frac{f^2}{2k}=-\frac{1}{2}k 
  \average{x}_f^2.
\end{equation}
This quantity is equal to the one obtained from the partition function,
as can be trivially checked. It is negative, because the positive variation
of the spring potential energy is more than compensated by the drop in
the potential energy of the applied force. One can imagine, e.g., that
the guide to which is constrained the particle is made rotated by
an angle $\theta$ around the origin in a vertical plane. Then 
$f=mg\sin \theta$, and, at equilibrium, the height of the particle
has dropped to $z=-\average{x}_f \sin\theta$, with a corresponding 
gravitational potential energy drop equal to $mgz$.

Of course nothing prevents us to take explicitly into account 
this contribution, in order to evaluate the change of $E-TS$, where
$E$ is the \textit{internal} energy. It is sufficient to subtract
the potential of the applied force. This is indeed what is routinely
made when estimating the free energy landscapes from manipulation
experiments, following the suggestions of Hummer and Szabo~\cite{HS}
and several others.

Let us now consider the second example discussed by Vilar and Rubí~\cite{VR},
namely a harmonically bound particle subject to
a harmonic time-dependent force:
\begin{equation}
  H(x,X)=\frac{1}{2}k x^2 +\frac{1}{2}K (x-X)^2.
\end{equation}
By expanding this expression, we obtain
\begin{equation}
  H(x,X)=\frac{1}{2}(k+K) x^2 -KX\,x+\frac{1}{2}K X^2.
\end{equation}
Then
\begin{equation}
  d W=-d X\,K (x-X).
\end{equation}
Of course
\begin{equation}
  d W_\mathrm{rev}=-d X\,K\left(\average{x}_X-X\right),
\end{equation}
where $\average{x}_X$ is the average displacement of the
particle when the center of the harmonic trap is placed
at $X$:
\begin{equation}
  \average{x}_X =\frac{K}{k+K}X.
\end{equation}
Thus
\begin{eqnarray}
  \Delta G&=&\int d W_\mathrm{rev}=
  \int_0^X d X'\;K \left(1-\frac{K}{k+K}\right)X'\nonumber\\
  &=&\frac{1}{2}\frac{k K}{k+K}X^2=
  \frac{1}{2}\frac{k(k+K)}{K}\average{x}_X^2.
\end{eqnarray}
This result is different from that reported by Vilar and Rubí, since
it takes into account also the change in  the potential energy $U$ of the
applied force. 

Let us evaluate the change of $\average{U}$. We have
\begin{eqnarray}
  \average{U}_0&=&\frac{1}{2}K \average{x^2}_0=\frac{1}{2}K 
  \average{\Delta x^2}_0;\\
  \average{U}_X&=&\frac{1}{2}K \average{\left(x-X\right)^2}_X\nonumber\\
  &=&\frac{1}{2}K \left[\left(\average{x}_X-X\right)^2 
    + \average{\Delta x^2}_X\right].
\end{eqnarray}
In our case, $\average{\Delta x^2}_0=\average{\Delta x^2}_X$. Thus
\begin{eqnarray}
  \average{U}_X-\average{U}_0&=&
  \frac{1}{2}K\left(\average{x}_X-X\right)^2 \nonumber\\
  &=&\frac{1}{2}K\left(1-\frac{k+K}{K}\right)\average{x}_X^2\nonumber\\
  &=&\frac{1}{2}\frac{k^2}{K}\average{x}_X^2.
\end{eqnarray}
We obtain therefore the difference $\Delta F$ of the \textit{internal} free
energy $F=E- TS=\left(\average{H}-U\right)-TS$:
\begin{equation}
  \Delta F=\Delta G-\Delta\average{U}=\frac{1}{2}k \average{x}_X^2.
\end{equation}
We see that, with the correct application of the thermodynamic relations,
it is possible to reconstruct the internal (free) energy landscape.

\begin{acknowledgments}
LP is grateful to Stan Leibler for pointing out to him ref.~\cite{VR}.
\end{acknowledgments}


\begin{thebibliography}{}
\bibitem{VR}
J. M. G. Vilar, J. M. Rubí, cond-mat arXiv:0704.0761v2.
\bibitem{Tolman}
R. C. Tolman, \textsl{The Principles of Statistical Mechanics} (Oxford:
Oxford U. P., 1938) (reprinted New York: Dover, 1979).
\bibitem{Jarzynski}
C. Jarzynski, Phys. Rev. Lett. {\bf 78}, 931 (1997).
\bibitem{HS}
G. Hummer, A. Szabo, Proc. Natl. Acad. Sci. USA {\bf 98}, 3658 (2001).
\end{thebibliography}
\end{document}